# Reanalyses of the Sunspot Observations of Fogelius and Siverus: Two "Long-Term" Observers during the Maunder Minimum

Hisashi Hayakawa (1–4)*, Tomoya Iju (5), Shoma Uneme (2), Bruno P. Besser (6, 7), Shunsuke Kosaka (1, 8), Shinsuke Imada (2)

(1) Institute for Advanced Researches, Nagoya University, Nagoya, 4648601, Japan

(2) Institute for Space-Earth Environmental Research, Nagoya University, Nagoya, 4648601, Japan

(3) UK Solar System Data Centre, Space Physics and Operations Division, RAL Space, Science and Technology Facilities Council, Rutherford Appleton Laboratory, Harwell Oxford, Didcot, Oxfordshire, OX11 0QX, UK

(4) Nishina Centre, Riken, Wako, 3510198, Japan.

(5) National Astronomical Observatory of Japan, 1818588, Mitaka, Japan.

(6) Space Research Institute, Austrian Academy of Sciences, Graz, 8042, Austria

(7) Institute of Physics, University of Graz, Universitätsplatz 5/II, Graz, 8010, Austria

(8) Graduate School of Humanities, Nagoya University, Nagoya, 4648601, Japan

* hisashi@nagoya-u.jp

**Abstract**

The solar activity during the Maunder Minimum (MM; 1645–1715) has been considered significantly different from the one captured in modern observations, in terms of sunspot group number and sunspot positions, whereas its actual amplitudes and distributions is still under active discussions. In its core period (1650/1660–1700), Martin Fogelius and Henrich Siverus have formed significant long-term series in the existing databases with numerous spotless days, as the 13th and 7th most active observers before the end of the MM. In this study, we have analysed their original archival records, revised their data, have removed significant contaminations of the apparent "spotless days" in the existing databases, and cast caveats on the potential underestimation of the solar-cycle amplitude in the core MM. Still, they reported at best one sunspot group throughout their observational period and confirm the significant suppressed the solar cycles during the MM, which is also supported from the contemporary observations of Hook and Willoughby. Based on the revised data, we have also derived positions of notable sunspot groups, which Siverus recorded in 1671 ($\approx$ N7.5° ± 2.5°), in comparison with those of Cassini's drawings ($\approx$ N10° ± 1°). Their coincidence in position and chronology in corrected dates indicates these sunspot groups were probably the same recurrent active region (AR) and its significantly long lifespan ($\geq$ 35 days) even during the MM.

**1. Introduction**

The MM is currently the only grand minimum captured in the instrumental sunspot observations since 1610 (Clette *et al*., 2014; Hathaway, 2015; Muñoz-Jaramillo and Vaquero, 2019; Arlt and Vaquero, 2020). This epoch is characterised by significant scarcity of sunspots, extremely asymmetric sunspot positions, and notably weakened solar magnetic activity (Spörer, 1889; Eddy, 1976; Ribes and Nesme-Ribes, 1993; Riley *et al*., 2015; Usoskin *et al*., 2015; Vaquero *et al*., 2015; Hayakawa *et al*., 2020c, 2020d), which is often explained with special dynamo activity and distinguished from normal cycle minima or even the Dalton Minimum (Usoskin *et al*., 2015; Hayakawa *et al*., 2020a, 2020b; Petrovay, 2020; Charbonneau, 2020). As the only grand minimum with telescopic solar observations, the MM has formed a reference for other grand minima detected in the proxy reconstructions from cosmogenic isotopes (*e.g.*, Usoskin *et al*., 2007, 2014; Inceoglu *et al*., 2015).

As such, it has been of great interest to analyse direct solar observations during the MM. Since the benchmark article of Eddy (1976), Hoyt and Schatten (1996) examined the observational coverage of major observers and incorporated continuous spotless days during this period. This result had formed a backbone for reconstruction of the sunspot group number during the MM in the existing database (Hoyt and Schatten, 1998 = HS98). Recent recalibrations of sunspot group number have required major revisions and updates for the exiting datasets including this period (Clette *et al*., 2014; Clette and Lefevre, 2016; Svalgaard and Schatten, 2016; Vaquero *et al*., 2016 = V+16). Apart from recoveries of forgotten records, recent studies have removed erroneous data and general descriptions on continuous spotless days during and around the MM (Carrasco *et al*., 2015, 2016, 2019; Vaquero *et al*., 2011, 2015, 2016; Usoskin *et al*., 2015; Arlt and Vaquero, 2020).

These efforts have improved our understanding on the MM and visualised its difference and similarity with other epochs. Despite its suppression, sunspot group number still shows extremely weak cycles with their core in 1660–1700 (Vaquero *et al*., 2015; Svalgaard and Schatten, 2016), while the ratios of umbra/penumbra look similar to the modern ones (Carrasco *et al*., 2018). Sunspot appearance was mostly concentrated in the southern solar hemisphere or near the solar equator, whereas this trend was probably not shared before its onset or its immediate afterwards (Carrasco *et al.*, 2019; Muñoz-Jaramillo and Vaquero, 2019). Nevertheless, the majority of the existing discussions are devoted to its latter half or recovery phase (see V+16; Muñoz-Jaramillo and Vaquero, 2019; Arlt and Vaquero, 2020). Therefore, analyses of the core MM (c.a., 1650/1660–1700) still improve our understanding on its depth (Vaquero *et al.*, 2015; Svalgaard and Schatten, 2016), which is currently an active discussion topic (Cliver and Ling, 2011; Lockwood and Owens, 2014; Owens *et al.*, 2017).

In this context, Martinus Fogelius' and Heinrich Siverus' observations have formed two long-term series interestingly located in 1661–1671 and 1671 – 1690, and were once used to depict the great deficit of sunspots in the 1660s and the early 1680s (Hoyt and Schatten, 1996; HS98). However, caveats must be noted on their almost continuous spotless days (V+16). In fact, Fogelius has been associated with a series of spotless days between 15 October 1661 and 31 July 1671 and Siverus with almost continuous spotless days with sporadic exceptions between 1675 and 1690. As such, V+16 casted doubts on their data validity in 1662–1670 among Fogelius', and that in 1675–1677, 1679, and 1681–1690 among Siverus'. Still, even removing these data, both Fogelius and Siverus have observational data for 318 days and 1040 days and hence have been considered the 13th and 7th most active observers before the end of the MM (V+16). Their observations are located around the beginning of the existing butterfly diagram for sunspot observations in the MM (Spörer, 1889; Ribes and Nesme-Ribes, 1993; Muñoz-Jaramillo and Vaquero, 2019) and will improve our understanding on the sunspot positions in this epoch.

Fogelius' original correspondence has been located in the Royal Society archives as a manuscript of MS EL/F1/34. This manuscript shows several unpublished details in Fogelius (1671), clarifying the original observer as Siverus and showing one more sunspot plate. Moreover, close inspections in Fogelius' correspondences and historical accounts for Siverus (Ettmüller, 1693) show actual availability of their datable observations. Therefore, in this article, we re-analyse the original sunspot records associated with Fogelius, reveal the actual observer, revise the sunspot group number and observational dates, derive sunspot positions, and compare them with contemporary observations.

## 2. Martin Fogelius' Sunspot Group Number

Fogelius' source documents are derived from Fogelius (1671) and Wolf (1850b), as clarified in the supplementary bibliography of HS98. Among them, Fogelius (1671) only involves a "plate showing course of spot from Aug 26 to Sep 5, 1671" (HS98). As such, majority of Fogelius's data are derived from Wolf (1850b, p. 295), whereas the spotless days are actually "general statements" (see Table II of Hoyt and Schatten, 1996) and Hoyt and Schatten (1998) even clarify, "we do not know exactly what days he was observing" (see HS98's bibliographical supplement). In fact, Wolf's original description is translated as follows: "Fogelius in Hamburg still has seen sunspots in October 1661. In 1671 Picard [NB Jean Picard] saw spots on 3 August and some following days, which were also observed in Hamburg on 7, 8, and 9 August and in Paris on 11, 12, and 13 August. On 30 August and 1 September, Hook saw one of these spots return. Siferus [NB Heinrich Siverus] in Hamburg followed it [NB the sunspot] continuously from 26 August to 4 September, and Fogelius communicated his observations on November 1st with a remark: '*Maculae solares iterum nobis apparuerunt, nunc tamen non amplius visibiles*' [NB indicating Fogelius (1671)] " (Wolf, 1850b, p. 295).

However, close inspection shows that the existing datasets for Fogelius and Siverus in HS98 and V+16 are not very robust considering their supplemental bibliography. First, the observations associated with Fogelius were actually those of Heinrich Siverus. Fogelius explicitly stated in his original letter (MS RS EL/F1/34) on 1 November 1671 in the Julian calendar and on 11 November 1671 in the Gregorian calendar; "Furthermore, the sunspot has revealed itself to us again but it is no longer visible. I send you the observations made here by the famous Heinrich Sivers [NB Siverus] with a telescope of modest length" (Hall and Hall, 1971, pp. 330–335). His statement is confirmed with signatures of "H. S." in the original drawings in the Royal Society archives (see Figure 1). On their basis, the sunspot observations in Fogelius (1671) are not by Fogelius himself but by Siverus.

Instead, Fogelius stated his own observations not in September but in August in his earlier letter (RS MS RS EL/F1/33) dated on 11 August 1671 in the Julian calendar (21 August 1671 in the Gregorian calendar). He stated, "this week we observed a sunspot here, demonstrated by the same very learned Picard, who had first seen it in the center of the sun's disk at sea, with a three-foot telescope, on the third/thirteenth of this month while off the island of Texel; in the following days it progressed towards the western limb near which we saw it on both the 7/17 and 8/18, each day nearer to the limb. On the ninth it was still visible and on the tenth we did not observe the heavens, because it was pretty cloudy. Whether Sivers saw it I may tell you next time. When Picard saw it the first time it was much like the tail of a scorpion, with a separated bulges which afterwards were observed to come together more closely. On the ninth it was like a melon seed" (Hall and Hall, 1971, pp. 198–201). On this basis, Fogelius' observational dates are identified as 17–19 August 1671, whereas the last one was overlooked in HS98.

Second, the general descriptions for spotless days require serious caveats for its usage on quantitative discussions (*e.g.*, Vaquero *et al.*, 2016; Arlt and Vaquero, 2020). Considering the modern weather of Hamburg, it is slightly unrealistic to expect either Fogelius to conduct observations every day during 15 October 1660 – 31 July 1671 as suggested in the existing databases (*e.g.*, HS98; V+16). A contemporary description in Philosophical Transactions makes us more cautious: "Whence it will appear, that some such Spots were seen here in London, A. 1660. And Mons. Picard affirm'd to Dr. Fogelius at Hamburg, that he had seen some in October 1661, witness the said Doctor's own letter, written to the Publisher August 11th last" (Anon., 1671, p. 2250). In fact, Fogelius himself explicitly stated, "Picard himself told us he has seen none [= no sunspots] since 13 and 14 October 1661, and had heard of none observed during the last ten years" in his correspondence (RS MS EL/F1/33; Hall and Hall, 1671, pp. 198–201). Accordingly, it was not Fogelius but Picard who saw sunspots on 13–14 October 1661 and affirmed a general zero statement by August 1671. Such descriptions do not provide clues for continuous spotless days for approximately ten years within Fogelius's own observations. Therefore, we need to remove Fogelius's data from the datasets of datable sunspot observations.

Fogelius' correspondence further shows his description on the lunar eclipse on 8 September 1671, which confirms that dating in his letter was based on the Julian calendar, as the total eclipse took place on 8 September not in the Gregorian calendar but in the Julian calendar. Accordingly, Siverus' observational dates should be converted to the Gregorian calendar. Moreover, another spotless day on 1 November 1671 (HS98; V+16) was not when Siverus or Fogelius conducted observations, but when Fogelius wrote his original letter (see Fogelius, 1671). Therefore, this spotless day should be removed.

### 3. Heinrich Siverus' Sunspot Group Number

As such, the original observer of the sunspot drawing in Fogelius (1671) has been revealed as Siverus, unlike what has been described in HS98. Archival investigations located his original letter in MS EL/F1/34 with two sunspot drawings, as shown in Figure 1. The one to the right is the original version of Figure III of Fogelius (1671). The other drawing involves three more observations: active days on 18 and 19 August in the Gregorian calendar and a spotless-day report on 20 August in the Gregorian calendar. This modifies Wolf's dating for his observations in early August in 1671.

Figure 1: Heinrich Siverus' original sunspot drawings in Fogelius's correspondence dated on 1

November 1671 (MS RS EL/F1/34; ©The Royal Society).

On the other hand, existing databases (HS98; V+16) have registered significant amounts of spotless days in Siverus' datasets in 1675–1690 (HS98) or at least in 1678 and 1680–1681 (V+16). Consulting the Wolf (1850a, pp. 46–47), the main source document for Siverus' observations is identified with Ettmüller (1693). Here, Ettmüller (1693, p. 5) has clarified that Siverus conducted sunspot observations from 1675 to 1690. Among this period, Ettmüller has specifically recollected sunspot visibility on 4 and 6 August 1680, 14 May and 15 June 1681 (Ettmüller, 1693, p. 8), 27–29 October 1689 (Ettmüller, 1693, p. 12), and 19–22 July 1689 (Ettmüller, 1693, p. 13). They are in Julian calendar and should be converted to Gregorian calendar (see Table 1). Their number has not been clarified that much, except for 27–29 October 1689, where the number of individual spots has developed from 3 to 6. Nevertheless, we can only derive visibility of sunspot groups on these dates and cannot be sure on its exact group. Therefore, we record their group number $\geq 1$, whereas those in 1689 had been removed in V+16.

On the other hand, neither Ettmüller (1693) nor Wolf (1850a, pp. 46–47) has documented spotless days with exact dates. In fact, HS98 has stated, "The original observations by Siverus are now lost so we do not know on what days he was observing. Spoerer (1889) [NB Spörer (1889)] says his manuscript in Hamburg could no longer be found". Therefore, Siverus' existing spotless days registered in the existing databases are without robust evidence and should be removed, unless otherwise his original logbooks are found in future.

Figure 2: (a) Sunspot drawings on "25–26 February 1678" depicted in Ettmüller (1693); (b) Willoughby's drawing for a spotless Sun on 22 June 1666 in Julian calendar, namely on 2 July 1666 (MS RS CLP 8i/13a; ©The Royal Society).

Apart from these records, Wolf mentioned, "From 25 and 26 February 1678, a large spot is depicted" in Ettmüller's dissertation (Wolf, 1850a, p. 46). This is probably derived from drawings in Ettmüller's dissertation (Figure 2a), which has been likely interpreted as Siverus' observations and incorporated in HS98 and V+16 after conversion to the Gregorian calendar (7 and 8 March 1678), while neither Ettmüller (1693) nor Wolf (1850a) has explicitly associated these drawings with Siverus. Therefore, we have removed them from Siverus' observations.

On their basis, datable sunspot observations in Fogelius' and Siverus' records have been summarized in Table 1, in comparison with his contemporaries' observations. As shown in Table 1, the apparent spotless days have been eliminated, as they were actually spotless days. Despite this drastic revision, Siverus' sunspot drawings provide valuable snapshots in the depth of the MM, showing only one group at best. The spotless days between 1662 and 1670 were actually misinterpretations of hearsays from Picard's observations. Still, Willoughby's report during the partial eclipse shows a spotless Sun on 2 July 1666 (MS RS CLP 8i/13a; see Figure 2b) and indicates a low solar activity in this interval. On the other hand, the only positive report in this interval, namely that of Atanasius Kircher on 2 September 1667 in V+16 is not an observational date but a date of his letter (Frick, 1681, p. 49; Usoskin *et al.*, 2015). Therefore, this record must be removed from the discussions on the sunspot group number, whereas this has been reserved in the revision of V+16 despite the clarification in Usoskin *et al.* (2015). Accordingly, the revised data show that the existing databases (*e.g.*, V+16) slightly underestimated the solar-cycle amplitude of the MM with these ghost spotless days, but that the solar activity in the MM is still much more suppressed than that of the modern solar cycles or the Dalton Minimum (see also Usoskin *et al.*, 2015; Vaquero *et al.*, 2015; Muñoz-Jaramillo and Vaquero, 2019; Hayakawa *et al.*, 2020a)

| Gregorian Calendar | | | Group Number | Reference | Observer |
|---|---|---|---|---|---|
| Year | Month | Date | | | |
| 1661 | 10 | 13 | 1 | MS RS EL/F1/33 | Picard |
| 1661 | 10 | 14 | 1 | MS RS EL/F1/33 | Picard |
| 1666 | 7 | 2 | 0 | MS RS CLP 8i/13a | Willoughby |
| 1671 | 8 | 13 | 1 | MS RS EL/F1/33 | Picard |
| 1671 | 8 | 17 | 1 | MS RS EL/F1/33 | Fogelius |
| 1671 | 8 | 18 | 1 | MS RS EL/F1/33 | Fogelius |
| 1671 | 8 | 18 | 1 | Fig. 1a | Siverus |
| 1671 | 8 | 19 | 1 | MS RS EL/F1/33 | Fogelius |
| 1671 | 8 | 19 | 1 | Fig. 1a | Siverus |
| 1671 | 8 | 20 | 0 | Fig. 1a | Siverus |
| 1671 | 9 | 5 | 1 | Fig. 1b | Siverus |
| 1671 | 9 | 6 | 1 | Fig. 1b | Siverus |
| 1671 | 9 | 7 | 1 | Fig. 1b | Siverus |
| 1671 | 9 | 8 | 1 | Fig. 1b | Siverus |
| 1671 | 9 | 9 | 1 | Fig. 1b | Siverus |
| 1671 | 9 | 9 | 1 | Hook 1671 | Hook |
| 1671 | 9 | 10 | 1 | Fig. 1b | Siverus |

| 1671 | 9 | 11 | 1 | Fig. 1b | Siverus |
|---|---|---|---|---|---|
| 1671 | 9 | 11 | 1 | Hook 1671 | Hook |
| 1671 | 9 | 12 | 1 | Fig. 1b | Siverus |
| 1671 | 9 | 13 | 1 | Fig. 1b | Siverus |
| 1671 | 9 | 14 | 1 | Fig. 1b | Siverus |
| 1671 | 9 | 15 | 1 | Fig. 1b | Siverus |
| 1676 | 8 | 8 | 2 | Hook 1677 | Hook |
| 1676 | 8 | 14 | 1 | Hook 1677 | Hook |
| 1676 | 8 | 17 | 0 | Hook 1677 | Hook |
| 1680 | 8 | 14 | $\geq 1$ | Ettmüller 1693 | Siverus |
| 1680 | 8 | 16 | $\geq 1$ | Ettmüller 1693 | Siverus |
| 1681 | 5 | 24 | $\geq 1$ | Ettmüller 1693 | Siverus |
| 1681 | 6 | 25 | $\geq 1$ | Ettmüller 1693 | Siverus |
| 1689 | 7 | 29 | $\geq 1$ | Ettmüller 1693 | Siverus |
| 1689 | 7 | 30 | $\geq 1$ | Ettmüller 1693 | Siverus |
| 1689 | 7 | 31 | $\geq 1$ | Ettmüller 1693 | Siverus |
| 1689 | 8 | 1 | $\geq 1$ | Ettmüller 1693 | Siverus |
| 1689 | 11 | 6 | $\geq 1$ | Ettmüller 1693 | Siverus |
| 1689 | 11 | 7 | $\geq 1$ | Ettmüller 1693 | Siverus |
| 1689 | 11 | 8 | $\geq 1$ | Ettmüller 1693 | Siverus |

Table 1: Fogelius' and Siverus' datable sunspot observations and those mentioned in Fogelius' correspondences and his contemporaries (Picard and Hook). Their original observers are corrected to Picard, Siverus, and Fogelius himself. Picard's observations on 3–4 August 1671 have been revised to 3 August 1671. We have also shown the revised version of Hook's sunspot group number for comparison.

## 4. Sunspot Positions and Comparisons with his Contemporary Observations

Obtaining corrected dates in the Gregorian calendar, sequences of Siverus' sunspot drawings in 1671 allow us to track motions and evolutions of this group, in combination with Cassini's oft-cited sunspot drawings in August 1671 (see Figure 2 from Cassini (1671a, 1671b); see also Usoskin *et al.* (2015)). Cassini made his observations at the Paris Observatory following the Gregorian calendar (*e.g.*, Ribes and Nesme-Ribes, 1993). Comparing Siverus' drawings (Figure 1) with Cassini's drawings, we infer that Siverus and Cassini probably reported the same sunspot group. Cassini had noticed this sunspot group earlier on 11 August (Figure 3a, II) and kept watching it until 19 August, when it almost arrived at the western limb (Figure 3b, VIII). Siverus monitored this sunspot group during the last two days: on 18–19 August (Figure 1a).

The August sunspot is probably identified with what Picard observed too (see Ribes and Nesme-Ribes, 1993). Cassini (1671a) cited Picard's observations as follows: "Mr. Picard had observ'd at Sea a Spot in the Sun from the third of August (ft. n.) to the nineteenth of the same inclusively; and seen it, at the first, like the Tayl of a Scorpion; but on the nineteenth day resembling a Melon-seed" (translated in Cassini (1671c, p. 2253)). Picard's extremely early witness on 3 August 1671 is actually a misinterpretation of the "thirteenth", as confirmed in Fogelius' letter (MS RS EL/F1/34; Hall and Hall, 1971, pp. 198–201) and his correspondence to Hevelius and Fogelius (Picolet, 1978, pp. 8 and 19). As such, his observations on 3–4 August 1671 should be removed from discussions of sunspot group number (*c.f.*, HS98; V+16). His description of its morphological transition from a scorpion tail to a melon seed looks consistent with Cassini's drawings (see Figure 3a (V) and 3b (VIII)).

Likewise, HS98 listed Stetini as reporting a sunspot group on 8–18 August 1671, citing Hevelius's *Machina Coelestis*. However, Hevelius himself stated "Macula ☉ notabilis conspecta Hamburgi a Cl. Picardo die 13. Stetini vero a die 8 as 18, quam etiam Lipsiae videreunt" which reads "A notable sunspot was seen at Hamburg by Cl. Picard on 13. In Stettin from 8 to 18, equally also seen in Leipzig" (Hevelius, 1679, p. 21). As such, "Stetini" is not a certain astronomer's name but just a name of the German city "Stettin".

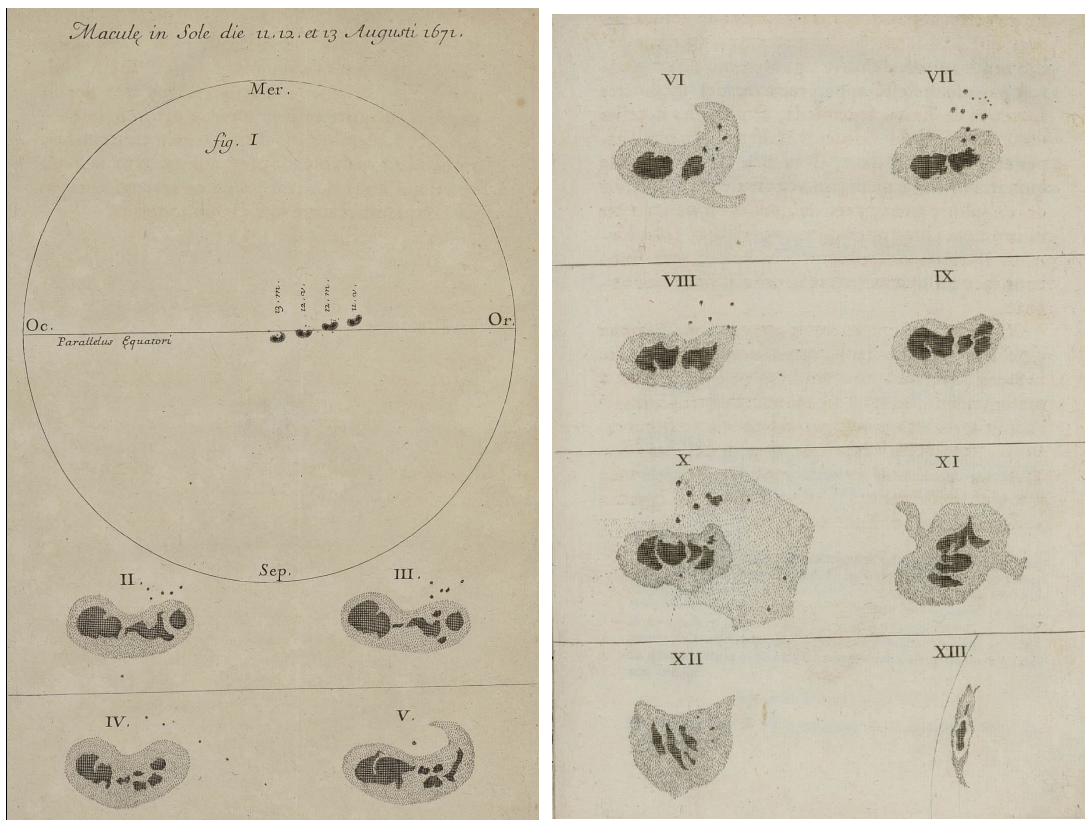

Figure 3: (a) Cassini's sunspot drawing with time sequence of a sunspot group in the whole-disk drawing during 11–13 August 1671 (I), with its enlargement on 11 (II), 12 (III and IV), and 13 August (V), adopted from Cassini (1671a); and (b) Cassini's enlarged sunspot drawings for the said group on 14 (I), 15 (II), 16 (III, IV, and V), 17 (VI), 18 (VII), and 19 August 1671(VIII), adapted from Cassini (1671b).

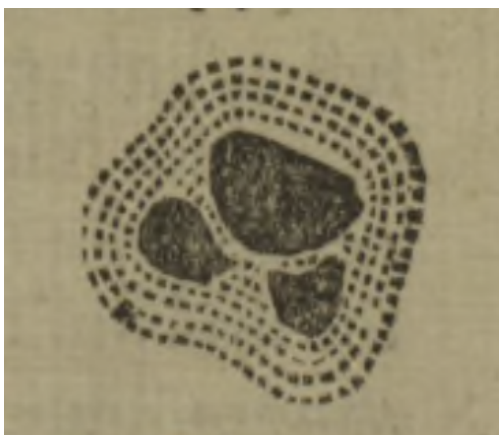

Figure 4: A close-up sunspot drawing at 15 LT on 11 September 1671, adopted from Hook (1671).

Cassini (1671a, 1671c) predicted the sunspot recurrence in September. This prediction was likely satisfied with Hook's observations in London, who "saw a large Spot in the Center of the Suns face about noon" on 30 August and the same group moving westward on 1 September in the Julian calendar (Hook, 1671, p. 2295), which should be corrected to 9 and 11 September in the Gregorian calendar with the drawing. Given their consistent central locations on 9 and 11 September, Hook probably saw what Siverus saw in September (see Figure 1b). This does not mean we should automatically convert all of Hook's observational dates in the existing databases (HS98; V+16). Three days of Robert Hook's observations are known in 1676 (25 June and 8 and 17 August). According to his original records, Robert Hook saw "a very conspicuous macula with its immediately incompassing Nubecula, and some other less conspicuous Spots at a further distance pass over the Disk of the Sun" in June 1676 without exact dates. He saw sunspots again "at the latter end of July and the beginning of August". On 29 July 1676 [NB 8 August 1676 in the Gregorian calendar], he saw "about six greater and smaller in one knot with their proper Nubecules or Umbra's" and lost this spot after 4 August [NB 14 August] and did not find any spots on 7 August [NB 17 August in the Gregorian calendar] (Hook, 1677, p. 52). Accordingly, we revise one record on 8 August 1676, remove one record on 25 June 1676, and add one record on 14 August 1676 in the existing datasets for Hook.

Relationships of these sunspot groups in mid-1671 are better assessed with the motion of these sunspot groups and comparison of their latitudes, as both Cassini (Figure 3a) and Siverus (Figures 1a and 1b) describe the motion of the sunspot groups in sequences. However, we need to be careful with their variable orientations. Siverus set his diagram orientation as east (*Or.*) at its left, west (*Occ.*) at its right, and north (*Sept.*) at its top, as clarified in his printed diagram (Fogelius, 1671, Figure III). However, Cassini set his diagram orientation as east (*Or.*) at its right, west (*Occ.*) at its left, south (*Mer.*) at its top, and north (*Sept.*) at its bottom, as shown in Figure 3a. The depicted directions indicate that Siverus' sunspot drawings are shown in erect images and Cassini's are shown in inverted images. Given these observations were performed in 1671, Siverus probably used a Schyrle-style refractor telescope with aerial imaging through filtering glass to show such erect images, whereas Cassini probably used a Kepler-style refractor telescope with aerial imaging through filtering glass to show such inverted images (see *e.g.*, Court and von Rohr, 1929; Rudd, 2007).

Following their date in the Gregorian calendar, we have tracked their motions. Both Siverus and Cassini have probably corrected the orientations of the solar disks for this comparison and show their North-South directions in the celestial coordinates rather than directions in the heliographic coordinates. On this basis, the solar rotational axis is found from the North-South direction on a drawing with a position angle (P-angle) correction. Using the corrected dates, the heliographic latitude of the centre of the visible solar disk (B0 angle) is computed as approximately 7°14′ for both the August (Figure 1a) and September drawings (Figure 1b). Taking these variables into account, we have computed the positions of depicted sunspots in Siverus' sunspot drawings (Figure 1) and Cassini's sunspot drawing (Figure 3a).

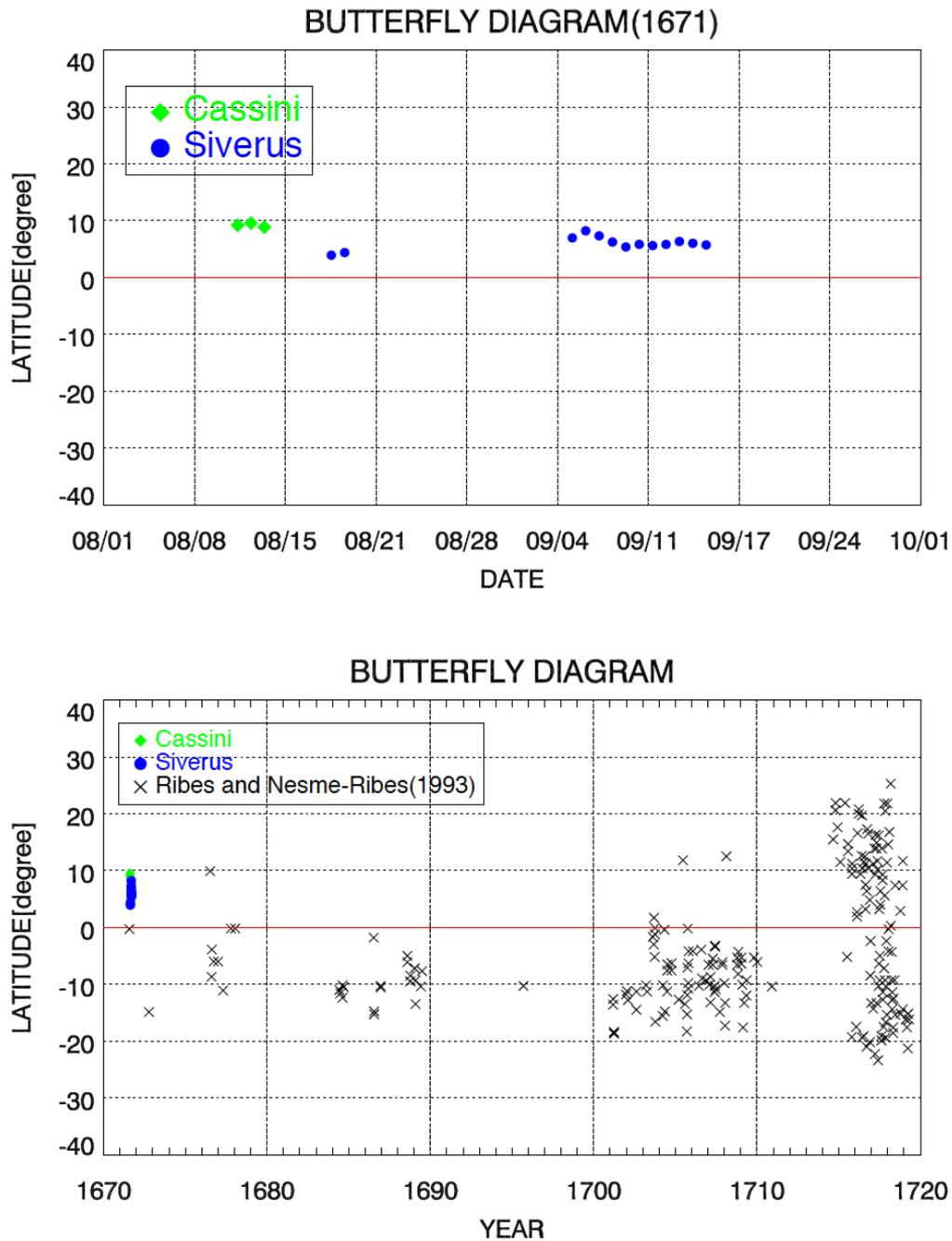

Figure 5: Positions of reported sunspots in Siverus' August and September drawings (Figure 1) in blue circles and Cassini's August drawing (Figure 3) in green diamonds above; their comparison with those in Ribes and Nesme-Ribes (1993) digitised in Vaquero *et al*. (2015b).

Figure 5 summarises these sunspot positions in our estimates. Our calculations show the sunspot moving in the latitude of $\approx$ N7.5° $\pm$ 2.5° in Siverus' August drawing and September drawing (Figure 1), except for the last three days (13–15 September), when the latitude of the sunspot monotonically increased. This anomalous excursion is probably caused due to its approach to the western solar limb. Therefore, we carefully recalculate positions of sunspots on September 13 – 15 to ensure that they are in latitudinal range above. Likewise, our calculations show Cassini's sunspot moving in the latitude of $\approx$ N10° $\pm$ 1° on 11–13 August. With similar latitudes, they were located near the disk centre on 13 August (see Figure 3) and 9 September (see Figure 1 and Table 1). Their 27-day separation agrees well with the duration of the solar rotation (*e.g*., Willis and Davis, 2015; Hayakawa *et al*., 2019). Therefore, their close latitudinal agreements and chronological separation for one solar rotation confirm that Siverus and Cassini probably depicted the same active region (AR) in the low latitude of the northern solar hemisphere and its recurrence in August and September to confirm speculations of Cassini and Hook.

## 6. Summary and Discussions

In this contribution, we have analysed the sunspot records, which have been associated with Fogelius in previous studies (HS98; V+16), in comparison with Fogelius' original correspondences (MS RS EL/F1/33 and MS RS EL/F1/34; Figure 1) and his contemporaries' observations such as those of Cassini and Hook (Figure 4). Our analyses have shown that the sunspot plate in Fogelius (1671) actually shows Heinrich Siverus' observations and have located original drawings including one omitted in Fogelius (1671).

Based on Fogelius' correspondences, we have shown that his and Siverus' records were dated in the Julian calendar and we have corrected the dates to the Gregorian calendar. Moreover, his sunspot observations on 13–14 October 1661 and few statements in the existing databases were not his own observations but actually hearsay from Picard. Fogelius' own observations are found on 17–19 August 1671. Siverus conducted his sunspot observations at least between 1675 and 1690, and a few of them have been cited in Ettmüller (1693) without explicit clarifications of exact dates of spotless days. We have revised Siverus' observations on their basis and removed apparent spotless days without robust evidence. These observations are so far the only known datable ones by Fogelius and Siverus, unless their original logbooks or separate publications are located. Therefore, the ghost spotless days in Fogelius' and Siverus' observations in the existing databases have probably underestimated the solar-cycle amplitude during the core MM.

In fact, their preserved data were not "long-term" observations but snapshots for the solar activity during the MM. Still, while these revisions removed considerable spotless days, they show at best 1 or 2 groups in this interval and do not contradict the absence of the positive days between 1662 and 1670. This is almost consistent with their contemporaries' observations such as those of Hook's and Willoughby's. Especially, between 1662 and 1670, Willoughby's recorded a spotless Sun on 2 July 1666 (MS RS CLP 8i/13a; see Figure 2b) and the only positive day in V+16, namely that of Atanasius Kircher on 2 September 1667 was not derived from his actual observation (see also Usoskin *et al*., 2015; c.f., Vaquero *et al*., 2016). These results have been summarised in Table 1 and indicate significantly low solar activity in this interval. As such, our result supports the current paradigm for the MM as a grand minimum (Usoskin *et al*., 2015; Vaquero *et al*., 2015) and disagrees with the claimed enhanced solar activity during the MM (Zolotova and Ponyavin, 2015).

Based on the revised data (see Table 1), we have compared their observations with Cassini's drawings in 1671, clarifying their orientations as erect images for Siverus' and inverted images for Cassini's. On this basis, we have computed their positions as approximately N7.5° ± 2.5° in Siverus' and approximately N10° ± 1° in Cassini's (Figure 5). Their latitudinal agreement and 27-day separation of the central meridian passages allow us to identify these sunspot groups probably as the same recurrent AR. Picard and observers at Stettin and Leipzig probably saw this AR too. We have revised Picard's observation date to 13 August and corrected Stetini and Leipzig in HS98 as observers at Stettin and Leipzig, both witnessing this sunspot on 8–18 August. This implies the AR had a notably long lifespan of ≥ 35 days even during the MM or at least an active longitude during the MM. For the former, recurrent ARs are more typical with larger ARs (Petrovay and van Driel-Gesztelyi, 1997; Henwood *et al*., 2010; Namekata *et al*., 2019) and hence those under higher solar activity (Hathaway et al., 2002; Figures 7 and 8 of Hathaway, 2015). Otherwise, this could indicate an active longitude during the MM. Further case studies are needed to exploit behaviours of the recurrent ARs during the MM.

### Acknowledgment

We thank Louisiane Ferlier, Katherine Marshall, Ellen Embleton, and other archivists in the Royal Society archives, for providing access to the archival materials such as Fogelius' correspondences, Jean-Pierre Allizart and Valérie Godin for their cordial helps upon HH's archival surveys at Paris, Koji Murata for his helpful discussions in philological issues, and Leif Svalgaard for his encouragements on our study. This work was supported in part by JSPS Grant-in-Aids

JP15H05812, JP17J06954, JP20K20918, and JP20H05643, JSPS Overseas Challenge Program for Young Researchers, JSPS Overseas Challenge Program for Young Researchers, the 2020 YLC collaborating research fund, and the research grants for Mission Research on Sustainable Humanosphere from Research Institute for Sustainable Humanosphere (RISH) of Kyoto University and Young Leader Cultivation (YLC) program of Nagoya University. This work has been partly merited from participation to the International Space Science Institute (ISSI, Bern, Switzerland) via the International Team 417 "Recalibration of the Sunspot Number Series", which has been organised by Frédéric Clette and Mathew J. Owens.

**Data Availability**

The original sunspot records of Fogelius, Siverus, and Willoughby are preserved in the Royal Society archives.